\documentclass[%
 reprint,
 amsmath,amssymb,
 aps,
]{revtex4-2}

\usepackage{graphicx}
\usepackage{dcolumn}
\usepackage{bm}

\begin{document}

\preprint{APS/123-QED}

\title{High-efficiency interface between multi-mode and single-mode fibers }

\author{Oussama Korichi}
 \altaffiliation {Tampere University, Photonics Laboratory, Physics Unit, Tampere, FI-33014, Finland}
 \author{Markus Hiekkamäki}
 \altaffiliation{Tampere University, Photonics Laboratory, Physics Unit, Tampere, FI-33014, Finland}
 \author{Robert Fickler}%
  \email{robert.fickler@tuni.fi}
   \affiliation{Tampere University, Photonics Laboratory, Physics Unit, Tampere, FI-33014, Finland}

\begin{abstract}
Multi-mode fibers (MMFs) and single-mode fibers (SMFs) are widely used in optical communication networks. MMFs are the practical choice in terms of cost in applications that require short distances. Beyond that, SMFs are necessary because of the modal dispersion in MMFs.
Here, we present a method capable of interfacing an MMF with an SMF using a re-programmable multi-plane light conversion scheme (MPLC). We demonstrate that only 3-phase modulations are necessary to achieve MMF-to-SMF coupling efficiencies from 30\% to 70\% for MMF's with core diameters up to 200µm. 
We show how the obtained coupling efficiency can be recovered if the output field of the MMF changes entirely, e.g. through strong deformation of the fiber, by simple monitoring of the field.
Furthermore, we test the influence of the resolution of both essential devices (field reconstruction and MPLC) on coupling efficiencies. 
We find that commercially available devices with increased speed and efficiency, such as wavefront sensors and deformable mirrors, are sufficient for establishing an MMF to SMF interface which auto-corrects any decoupling in the kHz regime.
\end{abstract}

\maketitle


\section{\label{sec:level1}Introduction}

The significant demand for handling information in communication networks has grown intensively in the last decades. 
To match the needs, different optical elements with higher performance have been developed, notably devices with broader bandwidths and the possibility for long transmission distances. 
Improvements in optical elements and system performance have boosted constantly the optical communications industry, since the 1970s \cite{arumugam2001optical}.
Optical fibers play a major role in communication networks. 
They are used in backbone networks as a transmission medium and cover either short distances (inside buildings or data centers) or long distances (regional and intercontinental distances).
Multi-mode optical ﬁbers (MMF), i.e. fibers with a large core diameter, are commonly used for short distances. 
However, MMFs allow multiple propagation modes with different optical path lengths, leading to modal dispersion and the formation of a speckle field at the output. 
Additionally, they are characterized by higher attenuation than single-mode fibers \cite{peng2019handbook}.
Hence, to overcome the MMF's limitation over long distances, single-mode optical ﬁbers (SMFs) are used.
SMFs have a core diameter of less than 10 $\mu$m (depending on the wavelength of the optical signal), allowing only one spatial mode to propagate. 
Owing to its lower attenuation and lack of modal dispersion, the effective transmission distance increases. 
However, the high cost of the equipment required to use SMFs makes MMFs the more cost-effective choice for short distances.
As each of the two fiber types, SMFs and MMFs, provide their own unique advantages, they are both used in global networks.
However, due to the difference in their nature, it is challenging to efficiently couple light from an MMF to an SMF.
Therefore, a device for efficiently interfacing an MMF with an SMF would be beneficial for optical telecommunication.

Multiple devices, which could in principle be used for such a task, have been proposed to realize an interface between a distorted wavefront, e.g., due to atmospheric turbulence, and an SMF.
Such devices include adaptive optics \cite{rinaldi2021sensorless} and multichannel free-space optical (FSO) receivers \cite{geisler2016multi,arikawa2018performance,yang2017multi}. 
Alternatively, a recent work \cite{billault2021free} uses a static multi-plane light conversion (MPLC) module as a multichannel FSO receiver to decompose the field distorted by turbulence into different single modes after which the signal is recombined into one single mode fiber via photonic integrated circuits.\\
In this paper, we present a simple method capable of achieving an efficient interface between an MMF and SMF using an active multi-plane light conversion scheme \cite{labroille2014efficient,fontaine2019laguerre,hiekkamaki2019near,brandt2020high}.
We show that only 3 phase modulations, realized with a single computer-controlled spatial light modulator (SLM), are required to achieve MMF to SMF coupling efficiencies of around 70\%, 60\%, 50\%, and 30\%, using MMFs with core diameters of 8.2 µm, 25 µm, 50 µm, and 200 µm, respectively. 
In addition, our system is monitoring the field and updating the MPLC conversion in real-time, such that the obtained coupling efficiency remains stable even if the output field of the MMF changes entirely, for example, through significant fiber deformation.
We further study the influence of resolution of the devices on the coupling efficiencies, both in the field reconstruction, implemented via an interferogram, and the light field modulation of the MPLC.
We find that even for relatively low resolutions in the field reconstruction (30x30 pixels), high coupling efficiencies can be obtained. 
In contrast, decreasing the resolution of the phase masks used in the MPLC significantly decreases the coupling efficiencies.
However, for a resolution of 60x60 pixels, good coupling efficiencies can already be achieved, such that efficient and fast commercially available devices could be used, which will pave the way for efficient MMF-to-SMF coupling in the kHz-regime.
\section{Methods}
At first, we outline the method used for field reconstruction and how the transformations were implemented to achieve high-efficiency single-mode fiber coupling.
\subsection{Field reconstruction}
\begin{figure}[htb]
\centering
\includegraphics[scale=0.7]{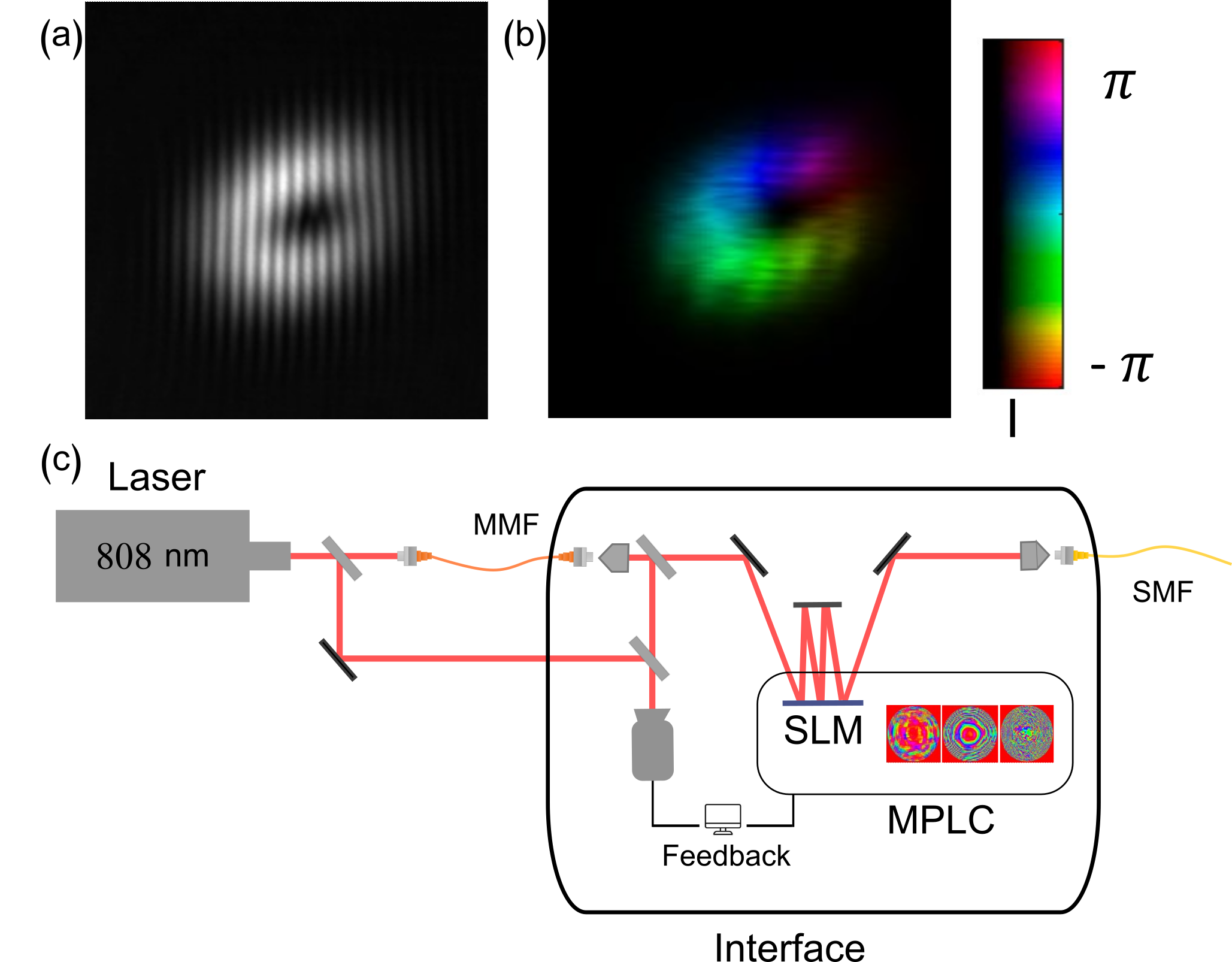}
\caption{Field reconstruction and sketch of the experimental setup to couple light from a multi-mode fiber (MMF) to a single-mode fiber (SMF). 
(a) An example of an interferogram between light coming from a few-mode fiber (8.2 µm core size) and the slightly inclined reference beam. 
(b) Reconstructed field  from the interferogram shown in a).
(c) Experimental setup to interface an MMF with an SMF using a multi-plane light conversion (MPLC) scheme implemented with a spatial light modulator (SLM). 
The MMF-to-SMF coupling is actively stabilized via a continuous monitoring of the complex amplitude of the output field of the MMF and a computer-controlled feedback for the MPLC scheme. }
\label{figure1}
\end{figure}

Standard cameras are only capable of accessing the transverse intensity structure of light lacking a direct way of measuring the complex amplitude, i.e. the transverse amplitude and phase distribution.
Although there are ways to directly obtain the full field information, e.g. a wavefront sensor, we decided to use an interferometric scheme \cite{takeda1982fourier} to increase the flexibility in choosing a wide range of different resolutions of the reconstructed field.
We generate an interferogram through interference of the field under investigation with a plane wave.
From the interference pattern, the complex amplitude can be extracted through simple fast Fourier transformation and appropriate filtering of the complex mode structure 
in the Fourier space. An example of an interferogram and the reconstructed field is shown in Fig. \ref{figure1}a and b.
\subsection{Field transformation}
Coupling light from an MMF to an SMF requires a transformation of the MMF field to a beam with a Gaussian profile.
We utilize an MPLC scheme to convert the measured MMF light field to the required SMF Gaussian field. 
A popular method to enhance coupling into SMFs is the so-called phase-flattening method which requires only a single-plane phase modulation \cite{qassim2014limitations}.
However, this method introduces considerable losses especially for complex beam structures such as speckle patterns out of large MMFs. 
In contrast, multiple phase modulation schemes enable, in principle, a lossless (unitary) transformation between an input and an output light field \cite{labroille2014efficient}. 
MPLC schemes have been applied in different settings from optical communication and multiplexing to distance estimation, atmospheric turbulence mitigation, complex beam shaping, and quantum operations \cite{boucher2017continuous,fontaine2019laguerre,hiekkamaki2019near,brandt2020high,boucher2020spatial, mounaix2020time, song2021simultaneous, hiekkamaki2021high,goel2022simultaneously}. 
The technique we used to obtain the required phase modulation screens is called wavefront matching (WFM), which is based on a waveguide design method \cite{hashimoto2005optical}. 
We apply the WFM method in our scheme to generate three phase modulations at three consecutive planes between which we allow for some free-space propagation.
More details and example code can be found in \cite{fontaine2019laguerre, hiekkamaki2019near, hiekkamaki_zenodo}.
In simulations, we found more than 97\% matching between the transformed field and a Gaussian.
\section{Experimental setup}
To experimentally investigate the coupling between various MMFs and an SMF, we implemented a setup, which automatically adjusts the MPLC modulation for best possible SMF coupling.
The experimental setup is illustrated in Fig. \ref{figure1}c. 
An 808 nm laser is split into two by a beam splitter. 
One half is magnified with a $4f$ system to have a quasi-plane wave, which is used as a reference beam to obtain the required interferogram for the reconstruction the complex field amplitude.
The second part of the beam is sent through an MMF of approximately 1m in length. 
To test the ability of the method for increasingly complex multi-mode fields, we test multiple fibers with core diameters of 8.2 µm, 25 µm, 50 µm, and 200 µm.
The MMF-to-SMF interface is realized by the aforementioned MPLC scheme implemented using a single SLM. 
The multi-mode beam from the output of the MMF is imaged on the SLM screen using a microscope objective.
In order to perform the correct modulation into an SMF field, the complex amplitude of the output field of the MMF needs to be reconstructed.
To enable the reconstruction via the earlier described interferometric scheme, we split a small part of the MMF field and interfere it with the reference beam.
After reconstructing the field, the phase modulations are computationally found with the WFM algorithm to transform the multi-mode field into the the SMF's eigenmode, i.e., a Gaussian beam profile. Each phase modulation covers an area of around 630 pixels by 630 pixels on the SLM screen, which has a pixel pitch of 8 µm along both of the axis. 
We achieve the mode conversion by displaying the three required holograms on separate parts of the SLM screen and bouncing the light field between the SLM and a mirror, which is placed 40 cm away. Owing to the limited efficiency of the SLM (75 \%), we display a grating structure for all phase modulations and only use the first diffraction order of each phase modulation.
After the MPLC scheme, the beam is directed to a  microscope objective (20x) and coupled to an SMF.\\
We note that due to imperfections of the SLM, the field undergoes some undesired distortions when modulated by the SLM. 
To compensate for these SLM-induced aberrations, we use a method based on the Gerchberg-Saxton algorithm \cite{jesacher2007wavefront} to counteract the effects and improve the beam quality. \\
The obtained corrective phase patterns were then permanently, displayed into the SLM.
Once the interface was setup, we continuously recorded the interferogram and updated the MPLC modulation such that our interface automatically corrected for beam disturbances in real time.
To test the capability of automatically correcting very strong beam deformations and keeping the high MMF-to-SMF coupling efficiency, we applied different fiber distortions.
In addition, we filtered for a different polarization of the output field of the MMF, using wave-plates and polarizers.
\begin{figure}[ht]
\centering
\includegraphics[scale=0.49]{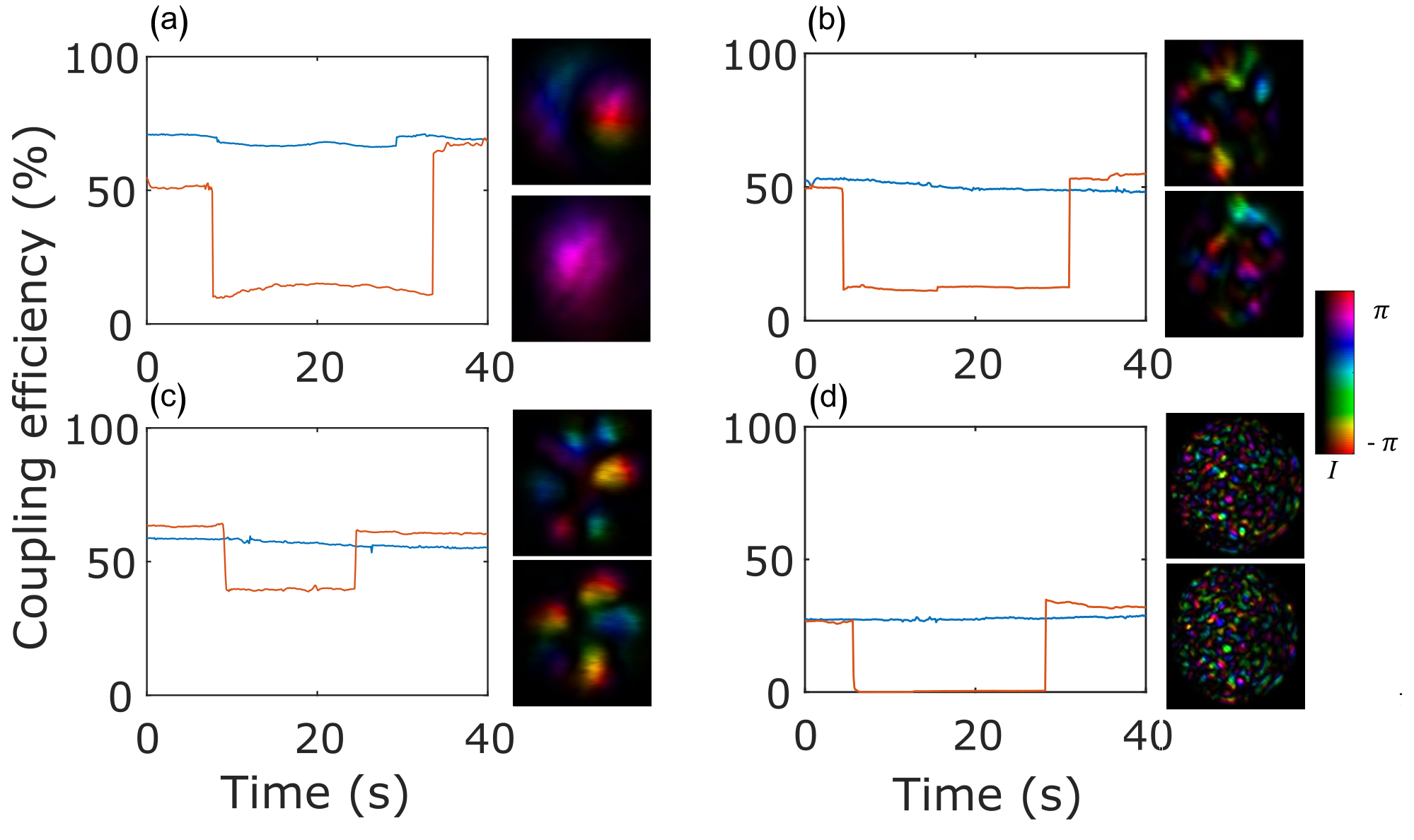}
\caption{Automatized coupling for strong MMF field deformations.
Stability of coupling efficiency without disturbance (blue) and automatized correction after a strong deformation of the fiber with core diameter of (a) 8.2 µm, (b) 50 µm, (c) 25 µm, and (d) 200 µm (orange). 
The insets on the right correspond to the complex field before (top) and after (bottom) the deformations for the different fields from the MMFs.}
\label{figure2}
\end{figure}
\section{Results}
Before testing our coupling scheme for MMFs, an SMF to SMF coupling is performed using the same setup to estimate the losses and imperfections in our system.
We measured a coupling efficiency of 85\% $\pm$ 3\% over 4 hours. 
The coupling efficiency is defined as $\eta=P_2/P_1$, where $P_1$ is the power at the end of the MPLC while $P_2$ is the power coupled into the SMF.
$P_1$ and $P_2$ are measured simultaneously using a beamsplitter with a known splitting ratio.
We attribute the measured losses to small misalignments and imperfect imaging in the setup, as well as some remaining imperfections in the light field's spatial mode caused by the SLM. 
\subsection{Multi-mode fibers}
At first, we test four different MMFs of different core sizes for which we achieve coupling efficiencies of 72\% $\pm$ 7\%, 60\% $\pm$ 3\% , 50\% $\pm$ 4\% , and 29\% $\pm$ 3\% over 4 hours of measurement.
The core diameters of the used MMFs were 8.2 µm, 25 µm, 50 µm, and 200 µm, respectively.
Here, as well as in the whole manuscript, the errors represent the standard deviation of the respective data set.
The decrease in coupling efficiencies results from the increasing complexity of the MMF field, as can be seen in the insets in Fig. \ref{figure2}. 
We attribute the varying errors to the challenges in a stable reconstruction of the correct field using the interferograms and the instabilities in alignments of the holograms within the MPLC system, which require a precision down to a few pixel. 
However, although the MMF fields significantly increase in complexity when increasing the MMF core size, our method is still able to achieve high coupling efficiencies.
\subsection{Automatic correction}
We deform the multi-mode field for testing the automatization of field reconstruction and mode transformation in real-time.
We found that when filtering for a orthogonal polarization at the output of the MMF, we achieved the biggest change such that the calculated overlap between the reconstructed fields before and after the polarization change was lower than 0.2\% for all fibers.\\
After around 3 seconds of a renewed field reconstruction and 22 sec of adapted phase modulation of the MPLC, we see that the initial coupling efficiency is re-established with only small variations. 
In total, we found that our setup requires around 25 seconds to restore the coupling efficiency, which is mainly limited by the slow field reconstruction, our non-optimized implementation of the WFM algorithm, as well as to some minor extent by the limited SLM refresh rate.
In Fig. \ref{figure2}, we show some examples of the achieved coupling efficiencies over time with and without deformations using the automatic re-coupling. 
We estimate that, by working with advanced commercial devices such as deformable mirrors and wavefront sensors in addition to implementing a faster version of the algorithm, stabilization times in the order of milliseconds or less can be achieved.
\subsection{Resolution}
\begin{figure}[h]
\centering
\includegraphics[scale=0.38]{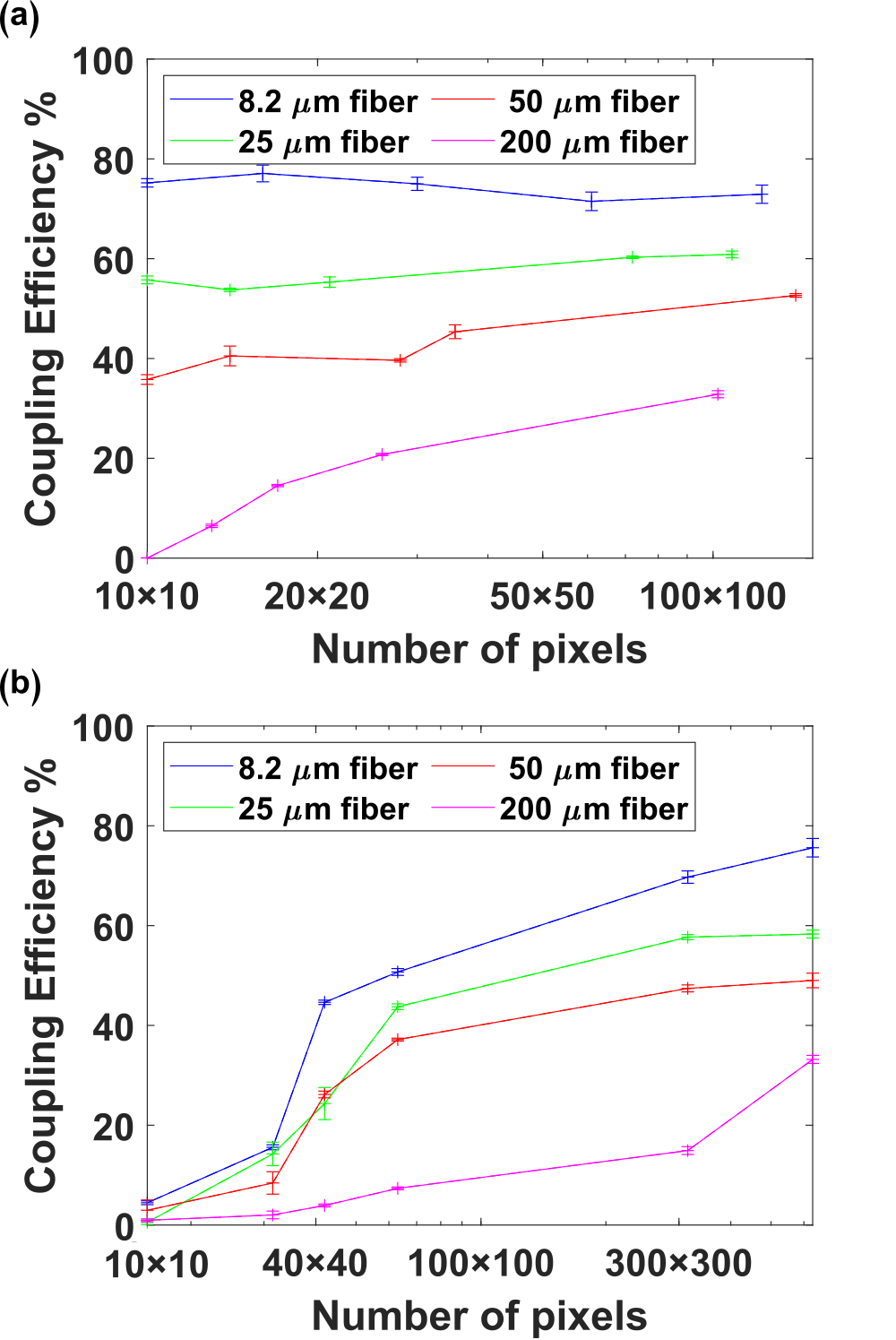}
\caption{SMF coupling efficiencies for different resolutions and MMFs depending on the field reconstruction (a) and phase modulations of the MPLC (b).
All efficiencies were obtained from 200 measurements within a time frame of 35 s and are illustrated as data points in a semi-log plot.}
\label{figure3}
\end{figure}

We additionally tested the effect of the device resolutions, both in the field reconstruction and MPLC, on the coupling efficiencies to gauge the limits for devices with lower resolution but higher refresh rate. \\
When testing the requirement in the field reconstruction for all studied fiber core sizes, we changed the resolution of the reconstructed mode from up to 140 x 140 pixels to minimally 10 x 10 pixels (see Fig. \ref{figure3}a) through averaging neighbouring pixels to obtain larger macro pixels.
For the two smallest fiber cores, i.e. 8.2 µm and 25 µm, we did not find a significant change in coupling efficiencies.
For a fiber core diameter of 50 µm, we found a small decrease from the above described maximal 50 \% to slightly less than 40 \% when using only 10 x 10 pixels.
For an MMF with a core diameter of 200 µm, the increased complexity of the MMF field led to stronger decrease in coupling efficiency, especially when the resolution gets smaller than 20 x 20 pixels, with complete decoupling for the the lowest resolution. 
Nevertheless, for all fibers, we found that a field reconstruction spanning 30 x 30 pixels was enough to achieve good coupling efficiencies such that higher resolutions are not necessary in principle.
Thus, commercially available cost-effective wavefront sensor with a resolution of 35 x 35 pixels would enable a high-efficiency coupling and automatized adjustment within tens of milliseconds and also lift the requirement for an interferometric measurement apparatus.
As a final set of measurements, we reduce the number of pixels used to display the phase holograms in the MPLC.
Here, we performed the WFM in the highest resolution at first and subsequently decreased the resolution of the found patterns through combining neighbouring pixels to one macro pixel with the average value.
We studied resolutions from maximally 630 x 630 pixels to only 10 x 10 pixels. 
In contrast to the resolution measurements in the field reconstruction, we found a rapid decrease for tested MMF core sizes when going to smaller resolutions, as can be seen in Fig. \ref{figure3}b.
For the fiber core diameters of 8.2 µm, 25 µm and 50 µm, a slow decrease can be witnessed with a steep decline at a resolution of around 30 to 60 pixels.
For the largest core size, already a change to 320 x 320 pixels led to a significant decrease in coupling efficiency with nearly zero efficiency reached at around 30 pixels. 
Again, we attribute the strong decrease to the vastly increased complexity of the MMF output field.
Hence, for MMFs with 8.2 µm, 25 µm and 50µm core diameters, standard deformable mirrors with a 60 x 60 actuator array could be used and still lead to high coupling efficiencies.
Such modulators easily achieve kHz refresh rates such that the MMF-to-SMF interface could be automated with high speed.
\section{Discussion and Conclusion}
Despite being very efficient, the presented MMF-to-SMF interface has its limitations, especially in terms of the time required to update the interface keeping the high coupling efficiencies.
The main tasks to improve will be to speed up the reconstruction of the field, by using a commercial wavefront sensor (e.g. Thorlabs WFS20r) or a faster implementation of the digital holography method \cite{carpenter2022digholo}, the WFM computation by implementing a faster version (e.g. using better hardware, and optimizing the computational efficiency), and a faster refresh rate of the modulation device such as deformable mirrors (e.g. Boston Micromachine Corporation DM 3K). 
In addition to a faster update of the interface, a reduction in resolution of the devices will also  simplify the alignment task due to the small number of pixels compared to our system.
As such, the proposed modifications might also increase the already good overall stability of the interface.\\
In summary, we presented a fully automated, high efficiency interface between MMFs of various core sizes and an SMF using an MPLC scheme with only 3 phase modulations.
We further showed that the spatial resolution of commercially available devices for field reconstruction and mode transformation, which offer simplified use, higher speeds, and increased overall efficiency, is good enough to maintain high coupling efficiencies. 
Thus, the proposed method can find applications in optical telecommunication and may also be adapted to correct for atmospheric disturbances in long-distance free-space communications \cite{dikmelik2005fiber}.
\section*{Acknowledgment} 
The authors thank Fr\'ed\'eric Bouchard for fruitful discussions.
The authors acknowledge the Academy of Finland (AKA) for support (PREIN - decision 320165). OK further acknowledges AKA (decision 336375), MH acknowledges the Doctoral School of Tampere University and the Magnus Ehrnrooth foundation, and RF acknowledges AKA through the Academy Research Fellowship (Decision 332399).
\section*{Disclosures} 
The authors declare no conflicts of interest.
\section*{Data availability}
Data underlying the results presented in this paper are
not publicly available at this time but may be obtained from the authors upon
reasonable request.
\bibliography{apssamp}

\end{document}